# BCI decoder performance comparison of an LSTM recurrent neural network and a Kalman filter in retrospective simulation

Tommy Hosman, Marco Vilela, Daniel Milstein, Jessica N. Kelemen, David M. Brandman, Leigh R. Hochberg, John D. Simeral, *IEEE Senior Member*

*Abstract*— Intracortical brain computer interfaces (iBCIs) using linear Kalman decoders have enabled individuals with paralysis to control a computer cursor for continuous point-and-click typing on a virtual keyboard, browsing the internet, and using familiar tablet apps. However, further advances are needed to deliver iBCI-enabled cursor control approaching able-bodied performance. Motivated by recent evidence that nonlinear recurrent neural networks (RNNs) can provide higher performance iBCI cursor control in nonhuman primates (NHPs), we evaluated decoding of intended cursor velocity from human motor cortical signals using a long-short term memory (LSTM) RNN trained across multiple days of multi-electrode recordings. Running simulations with previously recorded intracortical signals from three BrainGate iBCI trial participants, we demonstrate an RNN that can substantially increase bits-per-second metric in a high-speed cursor-based target selection task as well as a challenging small-target high-accuracy task when compared to a Kalman decoder. These results indicate that RNN decoding applied to human intracortical signals could achieve substantial performance advances in continuous 2-D cursor control and motivate a real-time RNN implementation for online evaluation by individuals with tetraplegia.

## I. Introduction

Intracortical brain computer interfaces (iBCIs) have enabled individuals living with severe motor disability to achieve point and click control of a computer cursor using imagined or attempted movements of their paralyzed hand [1]–[3]. Toward clinical viability, human iBCI research has also demonstrated rapid decoder calibration [4], sustained robust neural decoding [5], [6] and increasingly fast and accurate cursor control [5]–[7] enabling online chat, internet browsing, and the use of apps on a consumer tablet [8], [9]. Nonetheless, human iBCIs have not yet provided cursor performance equivalent to able-bodied hand movements. Candidates for iBCIs, particularly individuals with tetraplegia or locked-in syndrome, have a strong interest in using BCIs with high communication rates [10], [11], motivating further research to improve BCI-enabled cursor performance. In clinical trials, iBCIs have generally applied linear Kalman decoders to translate multielectrode intracortical signals into commands for assistive devices [5], [6], [9]. Recent results using nonlinear decoders in preliminary work in humans and in non-human primates (NHPs) show promise for even higher performance [12]–[16]. Specifically, in [13] a nonlinear recurrent neural network (RNN) variant outperformed the Kalman in electrode dropping experiments and achieved an increase in performance relative to the Kalman decoder in head-to-head comparisons. Motivated by the cursor performance gains with nonlinear approaches, we developed RNN methods for decoding human motor cortical signals into 2-dimensional cursor kinematics and evaluated performance relative to a steady-state Kalman decoder in simulated target selection tasks.

## II. Methods

This study evaluated the relative performance of a Kalman filter and a long-short term memory (LSTM) RNN in achieving fast and accurate cursor movements when decoding human intracortical neural signals recorded from implanted microelectrode arrays. The Institutional Review Boards of Partners HealthCare/Massachusetts General Hospital, Brown University, and Providence VA Medical Center granted permission for this study.

### A. Participants

This study was performed through simulations incorporating intracortical neural signals previously recorded from three participants with tetraplegia (T7, T9, T10) enrolled in a pilot clinical trial of the BrainGate* iBCI (ClinicalTrials.gov identifier: NCT00912041). Participants had two 96-channel microelectrode arrays (Blackrock Microsystems, Salt Lake City, UT) implanted [3] in the dominant hand area of the precentral gyrus except T10 whose second array was placed in the caudal middle frontal gyrus.

*Caution: Investigational device. Limited by federal law to investigational use. Research supported by Office of Research and Development, Rehab. R&D Service, Dept. of Veterans Affairs (N9288C, A2295R, B6453R, P1155R), NINDS (U01NS098968), NIH (S10 ODO1636 for cluster computation), ECOR of Massachusetts Gen. Hosp. (MGH), MGH-Deane Institute. The content is solely the responsibility of the authors and does not necessarily represent the official views of the NIH, the Department of Veterans Affairs or the US Government.

T. Hosman (thomas_hosman@brown.edu) and M. Vilela (marco_vilela@brown.edu) are with the School of Engineering and Carney Institute for Brain Science, Brown University, Providence, RI 02912.

D. Milstein (djmilstein@gmail.com) was with the Department of Computer Science and Carney Institute for Brain Science, Brown University.

J. Kelemen is with the Department of Neurology at Massachusetts General Hospital (MGH), Boston, MA.

D. Brandman was with the School of Engineering, Brown University. He is now with the Department of Surgery (Neurosurgery), Dalhousie University, Halifax, Canada.

L.R. Hochberg (leigh_hochberg@brown.edu) and J.D. Simeral (john_simeral@brown.edu) are with the VA Rehabilitation R&D Center for Neurorestoration and Neurotechnology, Dept. of VA Med. Ctr., Providence, RI, and the School of Engineering and Carney Institute for Brain Science at Brown University. L.R.H. is also with the Dept. of Neurology at MGH, Boston, MA and the Dept. of Neurology, Harvard Medical School, Boston.

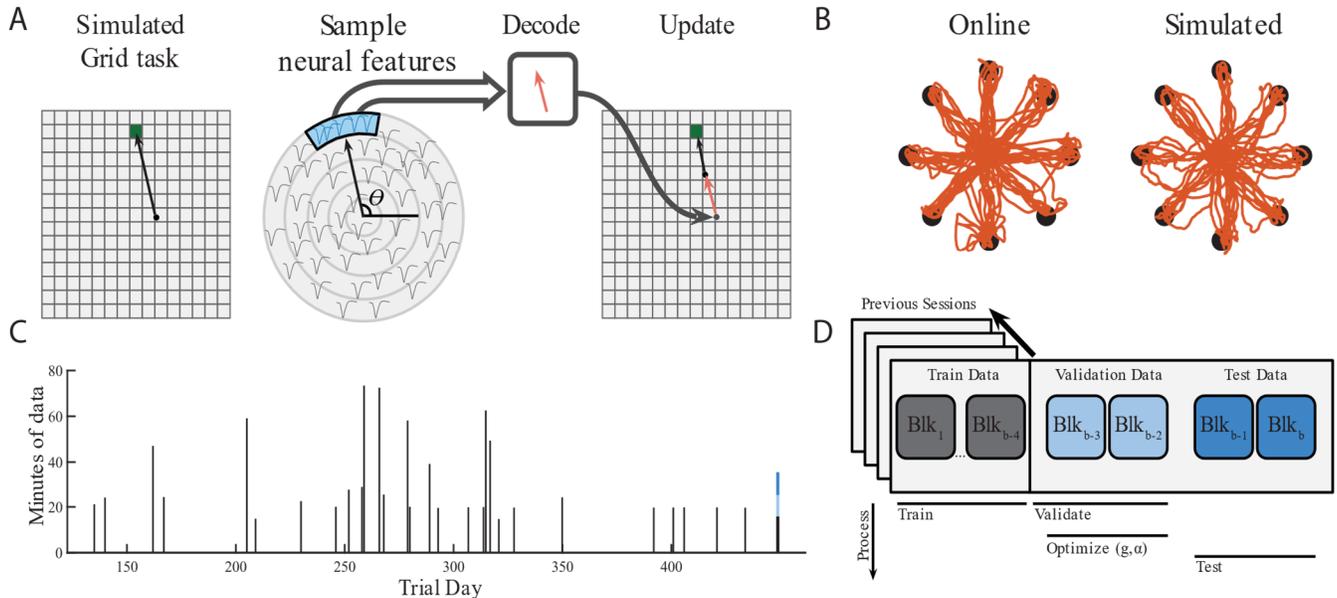

Fig. 1. (A) The process of simulating a Grid task with intracortical recordings. Given a cursor-to-target vector (black arrow), the simulator samples neural data with similarly labelled angle and distance-from-target. The decoder uses this data to make a predicted cursor update (red arrow). (B) Online and simulated cursor trajectories from a Radial-8 block. (C) A typical distribution of 30 sessions prior to a test sesion (rightmost line) demonstrating a typical amount of training data and and its distribution over days. (D) Data segmentation between training, validation and testing. Decoders were built using the training data, spanning D-days back from the test session. The trained decoders used validation blocks to optimize the gain and smoothing parameters for each decoder. Simulated comparisons were run from neural data sampled from the test blocks.

## B. Human Intracortical Neural Recording

Decoders in this study were trained and evaluated using intracortical microelectrode neural data originally recorded from participants in BrainGate research sessions in which neural features (multi-unit threshold-crossing spike rates and power in the spike-band LFP, 250 Hz – 5000 Hz) [4] were computed every 20 ms and decoded by a Kalman filter into instantaneous closed-loop cursor kinematics. In those original data sessions, participants used the Kalman iBCI to complete several closed-loop point-and-select cursor tasks: out-and-back Radial-4 or -8, Grid, and Random Target (mFitts) tasks [2], [4], [7], [17], [18]. Offline evaluation of alternative decoding algorithms was enabled by labeling (and saving) each 20 ms multielectrode neural feature with its corresponding online instantaneous 2-D cursor-to-target vector. The offline analyses looked at task blocks which lasted 3-5 minutes (a "block") consisting of at least 16 target acquisitions ("trials"). Neural features were z-scored per block and cursor-to-target vectors for each session were normalized by the 99$^{th}$ percentile of the session's labels. This scaling kept the majority of the 2D cursor-to-target vectors in the range of [±1, ±1], and prevented outliers from compressing the label ranges.

## C. Simulator

This study employed a custom simulator [19] that integrated a cursor task state machine, a decoding algorithm (either Kalman filter or RNN), and a process for selecting neural data samples from previously-recorded human intracortical data (Fig. 1A). For the purpose of performance comparisons, simulations executed a Grid task [6], [20] consisting of an n x n grid of adjacent square targets in which every square was selectable if the cursor remained on it for the specified selection period (dwell time). During the simulation, each instantaneous cursor position was translated into a distance and direction to the (simulator-generated) target (square on the grid). To determine the next-step cursor kinematics, the simulator drew neural features from a distribution of the participant's recorded 20 ms features whose originally-recorded labels had a similar cursor-to-target target vector (regardless of the absolute target position). The decoder-under-test used this neural data sample to compute a new velocity vector which was applied to move the cursor. The trial advanced in 20 ms time steps until the decoded cursor movements successfully acquired the target. Alternatively, an error trial resulted if the cursor dwelled on the wrong target or a timeout expired without a target acquisition; dwell and timeout durations differed with task as described below. The simulator then spawned the next target location and the task continued until 2 simulated minutes had elapsed. This procedure enabled offline decoding of participant-specific neural signals rather than artificially-created model-based data (e.g., cosine tuning). The sampling process enabled the generation of novel data-driven cursor trajectories responsive to instantaneous cursor and target positions.

## D. Decoder Assessment Overall Approach

The performance of Kalman and RNN decoders was evaluated on 86 different days (36, 27 and 23 "test sessions" for T7, T9 and T10, respectively). Test sessions were trial days with at least five blocks (total session blocks b >= 5).

For a given test session, neural data from all but the last two cursor task blocks {Blk$_{1:b-2}$} were used to calibrate Kalman and RNN decoders. Then a series of Grid task simulations was run sweeping parameter values (gain and smoothing) to maximize performance of each decoder on blocks {Blk$_{b-3:b-2}$} for that test session (details below). Finally, the performance of these optimized decoders was quantified in 2-minute Grid task simulations which drew neural "test data" from the test session's final two cursor task blocks {Blk$_{b-1:b}$} (Fig. 1).

Due to the stochastic nature of sampling from a pool of neural data at each step, every Grid task simulation was repeated 30 times to provide an average bitrate for that test configuration. Bitrate described the rate of communication in bits per second (bps) given N possible symbols [20]:

$$B = \frac{log_2(N-1)max(S_c - S_i, 0)}{t} \quad (1)$$

where $S_c$ and $S_i$ are the number of correct and incorrect targets acquired, respectively, and $t$ is the total time in the simulated task. This metric penalized incorrectly selected targets such that a selection accuracy of 50% results in a bitrate of 0 bps. We also measured target acquisition time as the average time the cursor took to move to and select a target, excluding incorrectly selected targets and timeout trials which were tallied as failed trials.

The entire foregoing procedure, including decoder training, optimization and evaluation, was repeated for each of the 86 test sessions.

*E. Speed and Accuracy Grid Task Variants*

The simulations described above assessed performance on two variants of the Grid task to evaluate how effectively each decoder could achieve both selection accuracy and cursor speed. A "high accuracy" Grid task required sustained selection of small targets (grid size n = 15, dwell time = 2 s, trial timeout = 10 s), thus penalizing noisy or imprecise cursor control. A "high speed" task presented larger grid targets with a shorter dwell requirement (grid size n = 10, dwell time = 0.5 s, trial timeout = 5 s), thus penalizing slow movements (through the risk of false selections) and requiring more rapid target selection to achieve a comparable bitrate. A high-performance BCI decoder should achieve a high combined bitrate across both speed and accuracy tasks. Therefore, each decoder was evaluated on both Grid task configurations using the same optimized gain and smoothing (see below).

*F. Decoder Parameter Optimization*

The choice of post-process parameter values has an important impact on decoder performance [21]. Both decoders applied a gain (g) to scale the decoder's output to the task workspace and determine the decoder's maximum speed. Additionally, the Kalman decoder had a parameter (α) that smoothed the decoder's output. As part of each session analysis described above, we swept these parameters in a series of Grid task simulations (g, 150 values for both the RNN and Kalman; α, 5 values for Kalman) to determine the test-session-specific parameter values that maximized the combined bitrates from the high accuracy and high performance Grid tasks. These simulations used the two pre-test blocks {Blk$_{b-3:b-2}$}, retaining the held-out test data blocks {Blk$_{b-1:b}$} for the final assessment simulations.

*G. Optimizing Training Set Size*

Different decoders benefit from different amounts of training data. Neural networks in particular may benefit from large datasets. To determine the effect of training set size on each decoder's bitrate performance, the high-speed and high-accuracy Grid tasks were simulated (including the optimization procedures above) with training on different numbers of prior sessions D {0:30}. The minimum D yielding the maximum average bitrate across Grid task types and participants was selected as the "optimal" D to train that decoder type for head-to-head comparison testing.

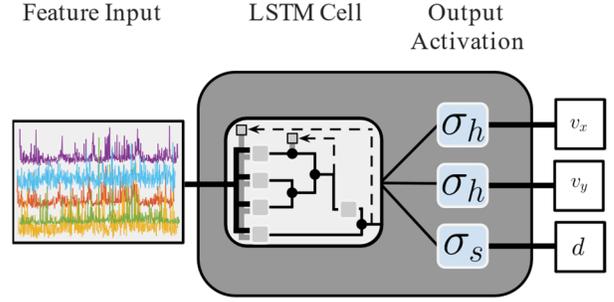

Fig 2. An RNN for decoding neural signals into cursor commands.

*H. Varying Grid Size*

To get a broader understanding of how each decoder's performance varied with task difficulty, we selected a dwell time (1 s) intermediate between the high speed and high accuracy tasks, a trial timeout of 5 seconds, and simulated the Grid task for a range of grid sizes {n = 2:25}. This sweep trained each decoder with its optimal days of data, D, and applied the foregoing simulation procedures to optimize gain and smoothing for each test session and grid size.

*I. Decoder Details*

*1) Kalman Decoder*

The steady state Kalman decoder used here is the same used by these participants in the BrainGate trial [5]. Briefly, this decoder fitted the state space matrix, $H$ (with ridge regression and 5-fold cross validation), based on the cosine tuning model. While the complete Kalman decoder continuously updates the Kalman gain matrix, $K$, it has been shown that $K$ can converge [1] which reduces the Kalman filter to a steady-state linear decoder of the form

$$v_t = g[Av_{t-1} + K[x_t - HAv_{t-1}]] \quad (2)$$

where $v_t$ is the decoded velocity, $x_t$ is the neural feature vector, $A$ is the identity matrix times a smoothing factor α, and $g$ is the post-process gain.

*2) Recurrent Neural Network (RNN)*

An RNN is a variant of a feedforward artificial neural network whose outputs feed back into the network's inputs [22]. This connection between output and input provides RNNs with implicit memory, which allows them to solve problems with temporal dependencies. For this study, we used a variant of the RNN with gated memory cells, the LSTM [23]. The basic LSTM cell with a forget gate is

$$f_t = \sigma_s(W_f x_t + U_f h_{t-1} + b_f) \quad (3)$$
$$i_t = \sigma_s(W_i x_t + U_i h_{t-1} + b_i) \quad (4)$$
$$o_t = \sigma_s(W_o x_t + U_o h_{t-1} + b_o) \quad (5)$$
$$c_u = \sigma_h(W_u x_t + U_u h_{t-1} + b_u) \quad (6)$$
$$c_t = f_t \circ c_{t-1} + i_t \circ c_u \quad (7)$$
$$h_t = o_t \circ \sigma_h(c_t) \quad (8)$$

where ∘ is the Hadamard product, $h_t$ is the LSTM cell's output, $c_u$ is the output from the cell update activation function and $c_t$ is the LSTM cell's internal state. $f_t$, $i_t$, $o_t$ are the output matrices from the respective forget, input, and output activation functions, which act as the network's gates and $W$, $U$, and $b$ represent the weights ($W$, $U$) and bias ($b$) for the respective activation functions ($f$, $i$, $o$, and $u$). $\sigma_s$ and $\sigma_h$ are the sigmoid and hyperbolic tangent functions.

The neural network architecture is simply an LSTM cell with three densely connected outputs (Fig. 2). Input features were passed directly to the RNN layer whose outputs went to three densely connected activation functions. Two decoded the x and y velocity $[v_x, v_y]$, and the third decoded the distance-to-target $d$, where the predicted distance was used to modulate the decoded direction vector, analogous to speed [24]. The post-process gain, $g$, further scaled the RNN's output to yield the RNN output velocity $v_t$:

$$v_x = \sigma_h(\boldsymbol{h}_t) \quad (9)$$
$$v_y = \sigma_h(\boldsymbol{h}_t) \quad (10)$$
$$d = \sigma_s(\boldsymbol{h}_t) \quad (11)$$
$$\boldsymbol{v}_t = gd[v_x, v_y] \quad (12)$$

The RNN was trained with the Keras library (Tensorflow backend) utilizing the Brown University computing cluster. The training hyperparameters used are listed in Table 1.

### III. RESULTS

#### A. Decoder Optimization

The RNN and Kalman decoders were trained with varying amounts of data to find the number of training sessions that maximized cursor bitrate on high speed and high accuracy Grid tasks averaged across participants (Fig. 3). For both tasks, maximum Kalman decoder performance (blue) was observed with training sets that included within-day data and one previous session. Beyond that, there was a general trend of declining performance with more training sessions. RNN performance (red) improved with seven to ten additional sessions of training data with limited or no further performance gain thereafter. After averaging across tasks to achieve decoders that balanced speed and accuracy (Fig. 3, bottom), the optimal Kalman filter training set included one session prior to the test sessions (59 ± 30 total training minutes, mean ± std). On average, D = 1 was 11 calendar days prior to the test session. The optimal RNN training set included seven prior sessions (250 ± 81 min) which on average spanned 73 calendar days (see Fig. 1C). These values of D were used for the results reported here.

TABLE I
Training Hyperparameters for the Recurrent Neural Network

| Parameter | Value | Parameter | Value |
| --- | --- | --- | --- |
| Hidden units | 50 | # features | 384 |
| Batch size | 512 | Optimizer | Adam |
| Learning rate | 0.001 | Dropout | 50% |
| Unrolled steps | 15 | Loss | Mean square error |

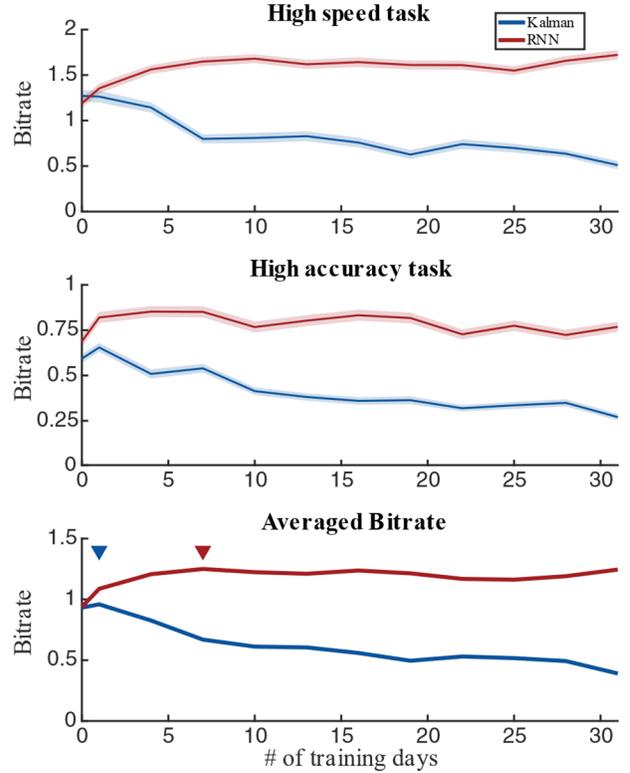

Fig. 3. Effects of training corpus size on bitrate for the Kalman and RNN. Average Kalman and RNN performance trained on a varying number of sessions for the high speed task (Top) and high accuracy task (Middle). Shaded region depicts standard error. (Bottom) The averaged bitrate from the high speed and high accuracy tasks. Arrows indicate peak average bitrate.

#### B. Cursor Performance Comparison

The Kalman decoder was compared to the RNN when decoding neural recordings from three participants. Within-session decoder comparisons used the optimal training corpus, gain and smoothing for each decoder. The decoders demonstrated both fast and accurate cursor movements by successfully completing both the high speed and high accuracy Grid tasks. Because we required a decoder to use the same set of session-optimized gain and smoothing values to complete both tasks, they sometimes failed to complete one or the other (a bitrate of 0 on all 30 runs of a given task). A test day was excluded from the comparison if either decoder could not complete either task. Excluded sessions consisted of 10 from just the RNN, 23 from just the Kalman, and 32 for both decoders. Direct bitrate comparisons for the remaining 21 test sessions are shown in Fig. 4A. The RNN achieved a higher bitrate than the Kalman in nearly 80% of the test sessions as demonstrated by the proportion of points above the diagonal. In many sessions, the RNN achieved bitrates in excess of 2.2 bps on the high speed task, a level never achieved by the Kalman in these sessions.

To more completely examine the range and consistency of decoder performance, Fig. 4B reports the bitrate observed in all 30 simulations for each test session (21 sessions x 30 simulations for each decoder and task). The RNN significantly outperformed the Kalman decoder on both tasks ($p < 0.001$ Wilcoxon rank-sum test, WRST) with a median bitrate on the high speed task of 2.5 bps for the RNN and 1.1 bps for the Kalman, and median bitrate on the high accuracy task of 1.2 bps for the RNN and 0.8 bps for the Kalman. On both tasks, the RNN frequently exceeded the best Kalman performance.

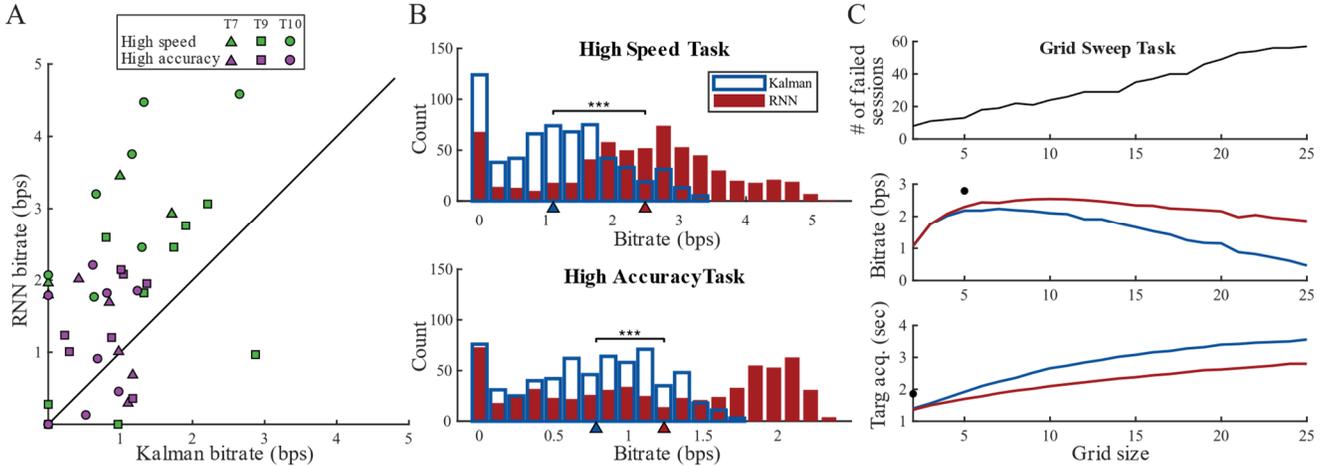

Fig. 4. Comparisons of RNN and Kalman decoder performance. (A) Median bitrate across 30 simulations for each test day for the high speed task (green) and the high accuracy task (purple). Points above the black diagonal line indicate test sessions with higher RNN performance, whereas points below indicate better Kalman performance. (B) Histograms of all simulation bitrates contributing to (A) for each task. Triangles indicate median Kalman and RNN bitrates (*** indicates $p < 0.001$, WRST). (C) Metrics from the grid size sweep study combined across all three participants. The number of sessions that yielded 0 bitrate (Grid task failure) for either the RNN or the Kalman increased as the grid became denser (top). Middle shows median bit rate for the RNN and Kalman as a function of grid size. Bottom shows the median target acquisition time as a function of grid size. Performance for each grid size in the lower two plots was computed excluding sessions tallied in the top plot. Standard errors for both lower plots could not be visibly plotted. For all grid sizes to the right of the dot, RNN performance was significantly better than the Kaman.

*C. Grid Size Sweep*

We found the performance of both decoders across a range of Grid task sizes (2 x 2 to 25 x 25, dwell = 1 s) for all 86 test sessions from the three participants (Fig. 4C). As the grid size increased (a more challenging task), fewer test sessions yielded viable cursor control (Fig. 4C, top) despite within-session gain and smoothing optimization. Eight test sessions failed to achieve even the simple 2 x 2 grid, consistent with poor overall in-session performance on those days. At the other extreme, both decoders successfully completed the most challenging Grid task on 27 of the 86 test sessions (31%). Bitrate performance of the two decoders (Fig. 4C, center) were not significantly different on the 2 x 2, 3 x 3 and 4 x 4 grids. The RNN significantly outperformed the Kalman on the more challenging tasks (n >= 5, $p < 0.001$ WRST) with a greater differential as the grid density increased. The RNN also significantly outperformed the Kalman across all grid sizes in terms of target acquisition time (Fig. 4C, bottom).

## IV. DISCUSSION AND FUTURE WORK

This study examined whether an RNN decoder could provide better two dimensional cursor performance than a Kalman filter used in recent iBCI research. Performance was evaluated across a range of decoder parameter values and Grid task difficulties using a simulator which sampled human intracortical signals previously recorded from three BrainGate trial participants as they completed blocks of closed-loop cursor control. For every session, each decoder was optimized by sweeping several parameters known to impact decoder performance, and then the performance of the optimized decoders was compared using hold-out data sets.

Across a wide range of task demands (grid sizes and dwell time requirements), an optimized RNN decoder consistently outperformed an optimized Kalman decoder in terms of bitrate and target acquisition time. The RNN decoders demonstrated higher performance in paradigms that required both high speed and high accuracy. The grid-sweep study demonstrated that the RNN consistently achieved higher bitrate and lower target acquisition time than the Kalman for grid sizes greater than 4, suggesting it may be better at enabling selection of small targets on busy cluttered computer desktops or complex application windows. The RNN performance gains were observed across three BrainGate trial participants. Together, these findings suggest that the RNN could be an effective decoder for high-performance clinical BCI use.

Human intracortical signals have been shown to exhibit nonstationarity over days [5], [25]. In part because of this, linear decoders are generally recalibrated using within-day neural data collected at the start of a BCI session [4], [5]. Consistent with this practice, we found near-optimal Kalman performance when calibration included only within-day data, with a slight average benefit when one prior session was also included [5]. Although RNNs generally benefit from high volumes of training data, nonstationarity made it unclear how much historical training data the RNN decoder would be able to utilize. On average, we found that peak RNN performance was achieved by training with data from 7 prior sessions, which here spanned 73 days on average.

Here, our model decoded both velocity and the cursor-distance-to-target (as an analog to speed). This approach appeared to stabilize the cursor at slow speeds (enabling sustained dwell periods) while allowing high post-process gain (enabling quick travel across the task workspace). While distance-to-target can correlate with motor task parameters such as movement amplitude or speed, it has been shown that distance-to-target and speed can be encoded separately in the neural activity of NHPs [26]. This motivates future work to test whether explicitly decoding speed could improve performance or if distance-to-target will be a sufficient speed analog for controlling BCI cursors online.

While the RNN exhibited significant performance advantages on both the high speed and high accuracy tasks, it did not always outperform the Kalman. One potential factor may be that a fast, responsive decoder such as the RNN might be more susceptible to even single-sample errors in the distance-to-target output during a dwell. This could cause the RNN cursor to prematurely jump off a small target and decrease the bitrate performance. Because performance depends critically on gain and smoothing parameters, we aimed to provide each decoder with its optimal gain (and, for the Kalman, smoothing) parameter values for each test session. Unlike the Kalman, the RNN architecture here only utilized intrinsic smoothing. Potentially, explicit smoothing heuristics and other optimizations may provide more robust decoding. Approaches for achieving optimal decoder gain and smoothing are the subject of ongoing research. Additional improvements may come from automatic feature extraction [27] and using feature embeddings [28] to map the relationship between the recorded intracortical neural activity and the intended movement.

Recurrent neural networks are computationally powerful tools which have been used in other studies to predict gait in NHPs [29], decode discrete hand movements in humans [30], and denoise neural data across several reaching and BCI tasks in NHPs and humans [15]. These studies, alongside this work, emphasize the potential of RNNs for BCI applications.

In this work, the RNN decoded 384 neural features in sub-millisecond time (median $0.6$ ms $\pm$ $0.1$ ms IQR) using an Intel i7 3.2 GHz processor with unoptimized code. This indicates that the RNN could be implemented in real-time for BCI applications. Although offline or simulated results may not always predict online cursor performance [31], these results motivate future investigation into the RNN's potential to enable cursor control for high bit-rate communication.

ACKNOWLEDGMENT

We thank BrainGate trial participants T7, T9, T10, their families and caregivers for contributions to this research.